\documentclass[prx,aps,showpacs,twocolumn,preprintnumbers,
amsmath,amssymb,superscriptaddress,longbibliography]{revtex4-1}
\usepackage[english]{babel}
\usepackage{amsmath,amssymb,amsfonts}
\usepackage{graphicx}
\usepackage[colorlinks=True,linkcolor=red,citecolor=blue,urlcolor=blue]{hyperref}
\usepackage{bookmark}
\usepackage{xcolor}
\usepackage{chngcntr}
\usepackage{braket}
\usepackage{nicefrac}
\usepackage{bm}
\usepackage{bbm}
\usepackage{braket}
\usepackage{slashed}
\usepackage[normalem]{ulem}
\usepackage{array}
\newcolumntype{C}{>{$}c<{$}}

\def\id{\mathbbm{1}}

\newcommand\scalemath[2]{\scalebox{#1}{\mbox{\ensuremath{\displaystyle #2}}}}
\allowdisplaybreaks

\begin{document}

\title{Suppression of second-harmonic generation from linear bands in the topological multifold semimetal RhSi}

\author{Baozhu Lu}
\thanks{These two authors contributed equally}
\affiliation{Department of Physics, Temple University, Philadelphia, PA 19122, USA}
 \author{Sharareh Sayyad}
\thanks{These two authors contributed equally}
 \affiliation{Univ. Grenoble Alpes, CNRS, Grenoble INP, Institut N\'eel, 38000 Grenoble, France}
\author{Miguel \'{A}ngel S\'{a}nchez-Mart\'{i}nez}
 \affiliation{Univ. Grenoble Alpes, CNRS, Grenoble INP, Institut N\'eel, 38000 Grenoble, France}
\author{Kaustuv Manna}
\affiliation{Max Planck Institute for Chemical Physics of Solids, Dresden D-01187, Germany}
\affiliation{Department of Physics, Indian Institute of Technology Delhi, New Delhi 110016, India}
\author{Claudia Felser}
\affiliation{Max Planck Institute for Chemical Physics of Solids, Dresden D-01187, Germany}
\author{Adolfo G. Grushin}
\email{adolfo.grushin@neel.cnrs.fr}
\affiliation{Univ. Grenoble Alpes, CNRS, Grenoble INP, Institut N\'eel, 38000 Grenoble, France}
\author{Darius H. Torchinsky}
\email{dtorchin@temple.edu}
\affiliation{Department of Physics, Temple University, Philadelphia, PA 19122, USA}

\begin{abstract}
Recent experiments in the topological Weyl semimetal TaAs have observed record-breaking second-harmonic generation, a non-linear optical response at $2\omega$ generated by an incoming light source at $\omega$.
However, whether second-harmonic generation is enhanced in topological semimetals in general is a challenging open question because their band structure entangles the contributions arising from trivial bands and topological band crossings. In this work, we circumvent this problem by studying RhSi, a chiral topological semimetal with a simple band structure with topological multifold fermions close to the Fermi energy. We measure second-harmonic generation~(SHG) in a wide frequency window, $\omega\in [0.27,1.5]$~eV and, using first principle calculations, we establish that, due to their linear dispersion, the contribution of multifold fermions to SHG is subdominant as compared with other regions in the Brillouin zone. Our calculations suggest that parts of the bands where the dispersion is relatively flat contribute significantly to SHG. As a whole, our results suggest avenues to enhance SHG responses.
\end{abstract}
\date{\today}
\maketitle

\paragraph*{\bf Introduction-.}
Second-harmonic generation~(SHG) is a nonlinear optical response that is useful in interrogating quantum phases of matter; since it only occurs in media without inversion symmetry, it is used as a proxy for spontaneous symmetry breaking~\cite{Shen86,Shen:1989vc,Dahn96,Petersen:2006uk,Harter295,Sirica2020} and in studies of the surface and interfacial properties of materials~\cite{Mizrahi1988,Pan1989,Chang1997,Kirilyuk2005,Hsieh2011,Lee2016}.
It is also widely applied technologically as the basis for generating light sources at different wavelengths~\cite{Robinson1973,Zhao2019}. Therefore, finding systems without inversion symmetry and with a high second-harmonic yield is a contemporary material science challenge.

A central challenge to finding materials with a large SHG is identifying the microscopic origin of large nonlinear optical responses.
In two recent experiments~\cite{Wu17,Patankar2018}, the topological semimetal TaAs~\cite{armitageRMP2018,WengPRX2015,LvPRX2015,Huang:2015vn,Yang:2015vi,XuScience2015,lvNatPhys2015} was reported to exhibit a giant SHG response at $\omega\sim$ 1.5~eV (800~nm)~\cite{Wu17}, reaching a maximum yield $\sim 2\times10^2$ larger than the maximum response of the semiconductor GaAs at $0.7$~eV incoming photon energy~\cite{Patankar2018}. The $\sim$ eV frequencies at which the band-structure was probed, however, were far larger than the $\sim 60$ meV energy scale associated with the topological degeneracies of its low-energy band structure, the Weyl nodes.
Hence, the existence of Weyl nodes cannot explain the enhanced response. Instead, the enhancement was attributed phenomenologically to the skewness of the polarization distribution~\cite{Patankar2018}, but a general microscopic origin has yet to be uncovered. Moreover, the role of topological degeneracies with linear dispersion, such as Weyl nodes, in determining SHG remains experimentally unclear, mainly due to the complex band structure of TaAs when probed at large ($\sim$eV) frequencies~\cite{Wu17,Li2018,Patankar2018}.

In this work, we show experimentally, and demonstrate theoretically, that transitions between linearly dispersing bands, specifically those close to topological band degeneracies, suppress rather than enhance SHG. We do so by studying the chiral topological semimetal RhSi in space group 198, which has a relatively simple band structure~\cite{changPRL2017,tangPRL2017,sanchez2019,Cochran2020} as compared with TaAs~\cite{WengPRX2015,Buckeridge16}. Close to the Fermi energy~($E_F$) three and four bands meet at the Brillouin center and corner, respectively, resulting in two topological degenerate points known as multifold nodes~\cite{Manes:2012fi,wiederPRL2016,bradlynScience2016,changNatMat2018,changNatMat2018}. 
Additionally, the cubic symmetry and the absence of inversion and mirror symmetries in space group 198 simplify the analysis of SHG from RhSi because, unlike TaAs, there is only one independent component of the SHG tensor, $\chi^{xyz}$. The simplicity of this space group has aided the interpretation of other non-linear optical responses, notably the circular-photogalvanic effect~\cite{rees2019,Ni2020a,Ni2020b}

\begin{figure}[t]
    \centering
    \includegraphics[width=\columnwidth]{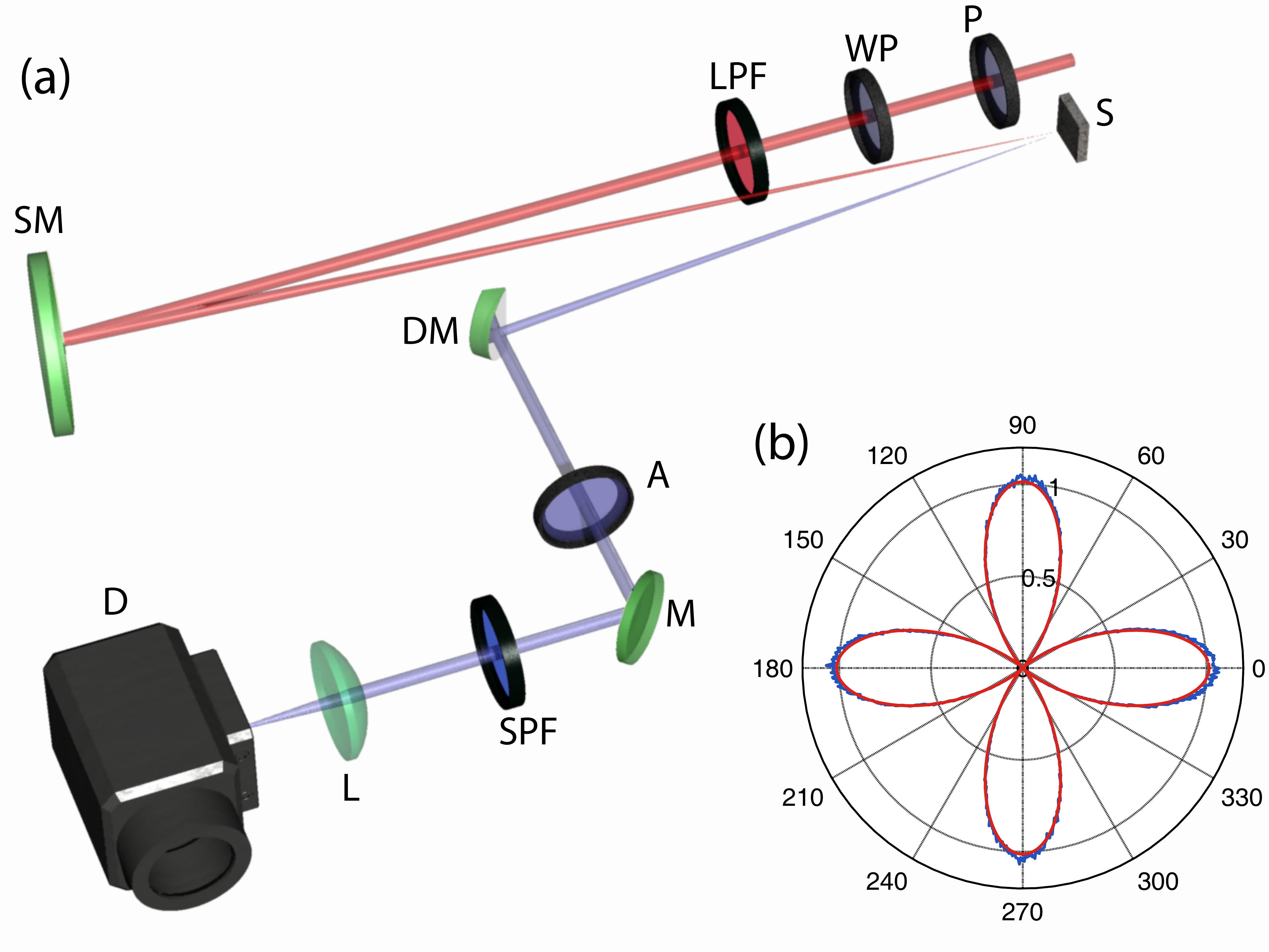}\\
    \includegraphics[width=\columnwidth]{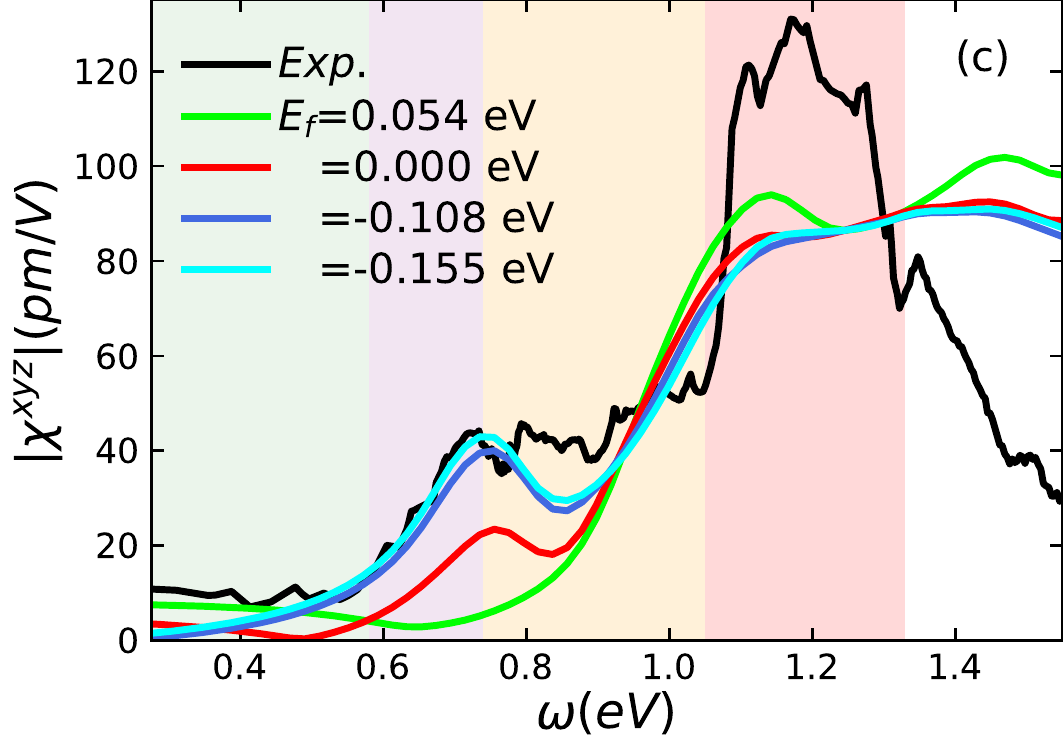}
    \caption{
    (a) Schematic diagram of setup used in SHG measurements. Optics are: P - polarizer, WP - waveplate, LPF - longpass filter, SM - spherical mirror, S - sample, DM - D-shaped mirror, A - analyzer, M - mirror, SPF - shortpass filter, L - lens, D - detector. (b) Representative data and fit for $\omega = 1.24$~eV. The data are in blue while the fit to Eq.~\eqref{eq:PP} is in red.
    (c) Experimentally measured~(black line) and theoretically calculated~(colored lines) SHG susceptibilities. Fermi energies are indicated by colors $E_f=$ 0.054~(green), $0.00$~(dark red), $-0.108$~(royal blue), $-0.155$~(cyan)~eV. The scissors potential is $\Delta=$ 1.23~eV. Shaded areas represent the photon energies at which different transitions from the valance to conduction bands occur. See also Fig.~\ref{fig:band}(a) for examples of these transitions. Shaded areas span $\omega \in [0.276,0.58]$~(green), $[0.58,0.74]$~(purple), $[0.74,1.05]$~(orange), and $[1.05,1.33]$~(red)~eV. 
    \label{fig:experiment_vs_theory}}
\end{figure}

We report $\chi^{xyz}$ of RhSi over a wide frequency range~(see Fig.~\ref{fig:experiment_vs_theory}), i.e., from $0.27$ to $1.55$~eV, and compare it with first principle calculations which, at low-energies, are also benchmarked with a $k\cdot p$ model~\cite{Ni2020b}. By identifying the regions in the band structure connected by optical transitions, we can infer that contributions between linearly dispersing bands are relatively small compared to those regions with relatively flat dispersion. When linear contributions are active (green and yellow regions in Fig.~\ref{fig:experiment_vs_theory}), the increase of the SHG signal as a function of frequency is relatively small compared to other frequency regions (purple and red regions in Fig.~\ref{fig:experiment_vs_theory}). The best agreement with the data is obtained after correcting the bare separation between bands by incorporating many-body effects~\cite{Levine1989, Hughes1996, Nastos2005, Sadhukhan2020, Song2020}, suggesting that capturing other nonlinear responses in chiral topological semimetals may require these corrections as well.  

At the single-particle level, the suppression of SHG due to linear bands can be understood from dimensional analysis~\cite{yangarXiv2017,Pozo2020}: since the SHG susceptibility $\chi$ has units of inverse energy squared (in units of fundamental constants) and the linear bands have no associated energy scale, the first finite contribution to SHG is due to quadratic corrections to the linear bands. This contribution is frequency independent because, by dimensional analysis, the SHG may scale as $1/t^2$ where $t$ is inversely related to the band curvature. Hence, linear bands, where $t$ is large, have smaller contributions than other points in the Brillouin zone. In contrast, flatter parts of the Brillouin zone contribute with a larger density of states, resulting in a comparatively larger SHG.

\paragraph*{\bf Experiment-.}
RhSi crystalizes in the cubic space group $P2_{1}3$~(number 198). Several materials in this space group, notably CoSi, RhSi, AlPt, PdGa and PtGa~\cite{tangPRL2017,changPRL2017, raoNature2019, sanchez2019,takanePRL2019,schroterNatPhys2019,Schroter179,Yao:2020cc,Sessi:2020dw} are known chiral topological semimetals that lack inversion and mirror symmetries~\cite{Manes:2012fi,wiederPRL2016,bradlynScience2016,changNatMat2018}. Photoemission experiments revealed that these materials showed spectra consistent with a three-fold degeneracy at the $\Gamma$ point and a four-fold degeneracy at the zone corner~\cite{sanchez2019, raoNature2019, takanePRL2019,schroterNatPhys2019,Schroter179}. These are topological band degeneracies and lead to exotic photogalvanic effects, including a quantized circular photogalvanic effect~\cite{dejuanNatComm2017,Konig2017,changPRL2017,flickerPRB2018}, which has been proven to be challenging to observe~\cite{rees2019,Ni2020a,Ni2020b}.

Figure~\ref{fig:experiment_vs_theory}(a) shows a schematic diagram of our SHG setup. The output of a regeneratively amplified Ti:sapph laser producing 1.2~mJ, 35~fs pulses centered at 800~nm at a repetition rate of 5~kHz was used to pump an optical parametric amplifier~(OPA) from which we derived the incoming fundamental laser field in the 800~nm - $4.5~\mu$m wavelength range ($0.276 - 1.55$~eV). More details on the experimental system can be found in the Supplemental Material~\cite{SuppMat}. The intensity of the vertically polarized SHG output was measured as a function of incoming polarization angle $\phi$, an example of which is shown in Fig.~\ref{fig:experiment_vs_theory}(b) with a typical fit to the expression $2/3\left[\chi^{xyz}\cos(2\phi)\right]^2$. The fits were corrected for the experimental parameters of pulse duration, spot size and instrument response, and then normalized against a GaAs standard in order to arrive at an absolute quantitative value for the SHG susceptibility element $\chi^{xyz}$, with results in a ratio $\chi^{xyz}_{\textrm{GaAs}}/ \chi^{xyz}_{\textrm{RhSi}} = 2.4$ for photon energy $\omega=1.24$~eV.  The resulting SHG in the $0.27-1.5$~eV energy range is shown in Fig~\ref{fig:experiment_vs_theory}(c).

\paragraph*{\bf Theory-.}

We have carried out the density functional theory~(DFT) calculation using the EXCITING package~\cite{Gulans2014}, based on state-of-the-art full-potential linearized augmented plane wave implementations. We have employed the generalized gradient approximations within the Perdew-Burke-Ernzerhof scheme~\cite{Perdew1996} as an exchange-correlation functional. The lattice parameters of the chiral cubic crystal RhSi have been chosen based on experimental measurements~\cite{Engstrom1965, Chang2017}. Four atoms of Rh and four atoms of Si in the unit cell are located in the Wyckoff positions for the space group $P2_{1}3$~\cite{Engstrom1965, sanchez2019}. We have performed our calculations on a $40 \times 40 \times 40$ k-point grid. As band splitting due to the spin-orbit coupling is of the order of meV~\cite{Chang2017}, much smaller than the reported scattering strength ($\delta \sim 100$ meV~\cite{rees2019,Ni2020a}), we have neglected this effect in our calculations.

\begin{figure}[t]
    \centering
    \includegraphics[width=\columnwidth]{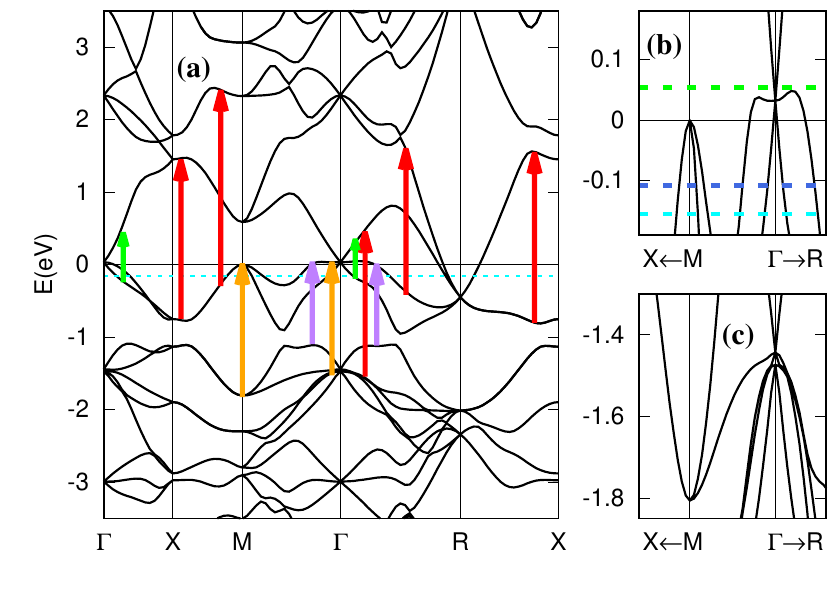}
    \caption{(a)~First-principles band structure of RhSi without spin-orbit coupling. Arrows indicate representative two-photon electronic transitions in SHG, and their color code corresponds to that of the shaded areas representing different frequency windows in Fig.~\ref{fig:experiment_vs_theory}(c). (b)~Zoom to the low energy bands between $\Gamma$ and $M$ points close to the Fermi level. Dashed lines indicate the Fermi energies at which the theoretical curves in Fig.~\ref{fig:experiment_vs_theory}(c) are plotted, namely $E_f=$ $0.054$~(green), $-0.108$~(royal blue), and $-0.155$~(cyan)~eV. (c)~Same as (b) but close to energy $-1.6$~eV. The zero of energy scale represents the Fermi energy of the pristine system.
    \label{fig:band}
    }
\end{figure}

The electronic band structure for RhSi along the lines connecting high symmetry points in the Brillouin zone is shown in Fig.~\ref{fig:band}. The energy is measured with respect to the Fermi energy of the pristine system $E_f=0$~eV. Close to the Fermi energy, the electronic structure possesses a threefold degeneracy at $\Gamma$ point, Fig.~\ref{fig:band}(b), and a fourfold degenerate point at $R$ point. We note that degenerate threefold crossings also exist at different energies at the $\Gamma$ point, e.g., around $E\approx-1.57$~eV, a region magnified in Fig.~\ref{fig:band}(c).

Our ab-initio results for the nonlinear susceptibility $\chi^{xyz}$ of RhSi are shown in Fig.~\ref{fig:experiment_vs_theory}(c), see Ref.~\cite{SuppMat} for more details. 
As in the experimental analysis, we also calibrate our results with GaAs~\cite{Bergfeld2003}. 
To account for the effects of disorder and finite temperature in our experimental sample, we have employed a Gaussian broadening with width $\delta=0.1$~eV, consistent with previous findings~\cite{Ni2020a}. We also include a scissors shift~\cite{Nastos2005} of $\Delta=$ 1.23~eV to
account for inaccurate band gaps between the occupied and unoccupied bands, see Ref.~\cite{SuppMat}. 

Our results for RhSi, for Fermi energies that lie above the threefold node~($E_f=0.054$, green line), and three that lie below this node~($0.0,-0.108,-0.155$~eV), are shown in Fig.~\ref{fig:experiment_vs_theory}(c). These spectra are similar in magnitude to other materials in the same space group and in the transition metal silicide family; we show the SHG spectra we computed for CoSi and MnSi in Ref.~\cite{SuppMat}. For RhSi, we observe that for $\omega \lesssim 0.45$~eV, theory and experiment agree well when $E_f = 0.054$~eV, while for $\omega > 0.45$~eV, it is the $E_f=-0.108,-0.155$~eV~(cyan and royal blue lines) curves that better reproduce the experimental data.

\begin{figure}
    \centering
    \includegraphics[width=\columnwidth]{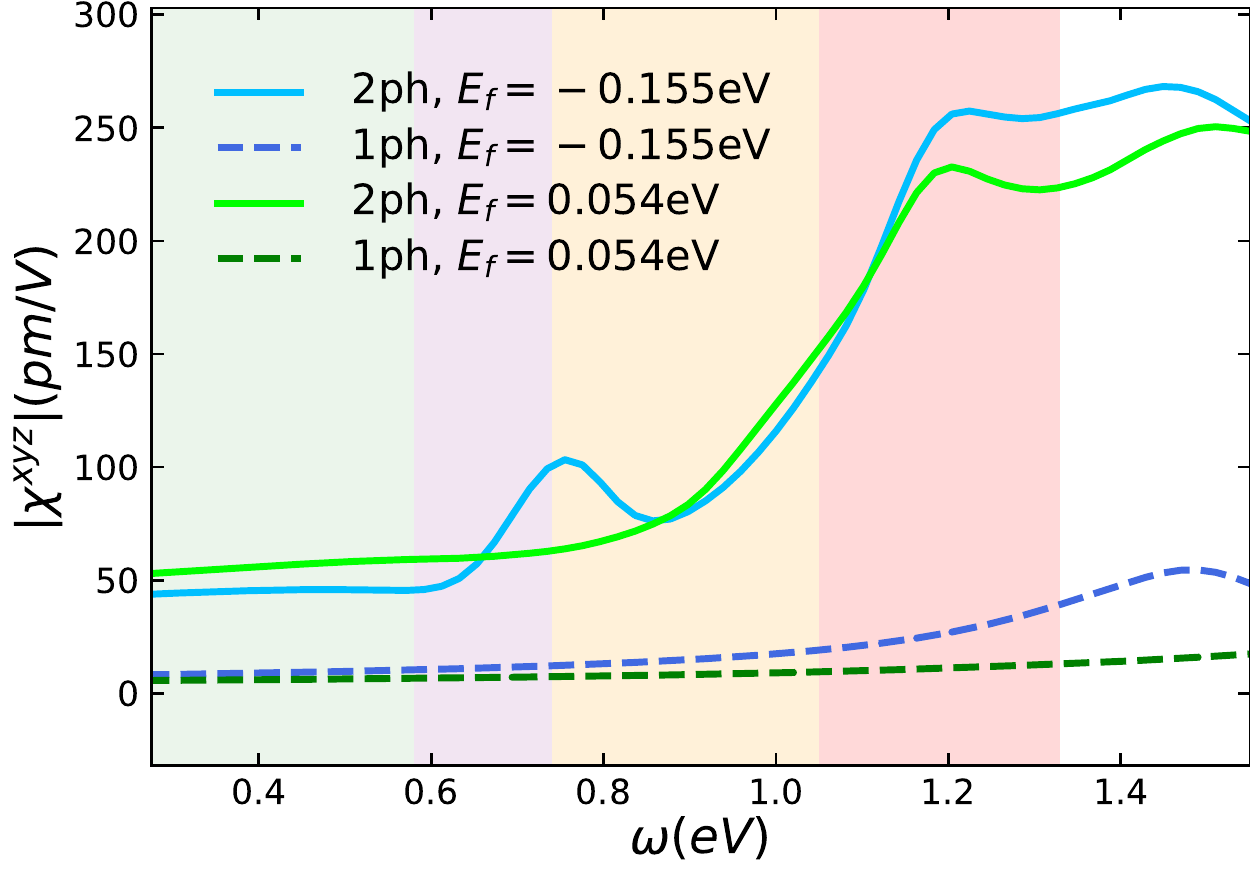}
    \caption{Calculated different components of SHG from two-photon~(solid sky blue line), one-photon~(dashed royal blue line) transitions with Fermi energy $E_f=-0.155$~eV as well as two-photon~(solid green line), one-photon~(solid dark green line) transitions with Fermi energy $E_f=0.054$~eV. The total SHG susceptibility is plotted in Fig.~\ref{fig:experiment_vs_theory}(c) and scissors potential $\Delta=1.23$eV.
    \label{fig:components_chi} 
    }
\end{figure}

\begin{figure}
    \centering
    \includegraphics[width=\columnwidth]{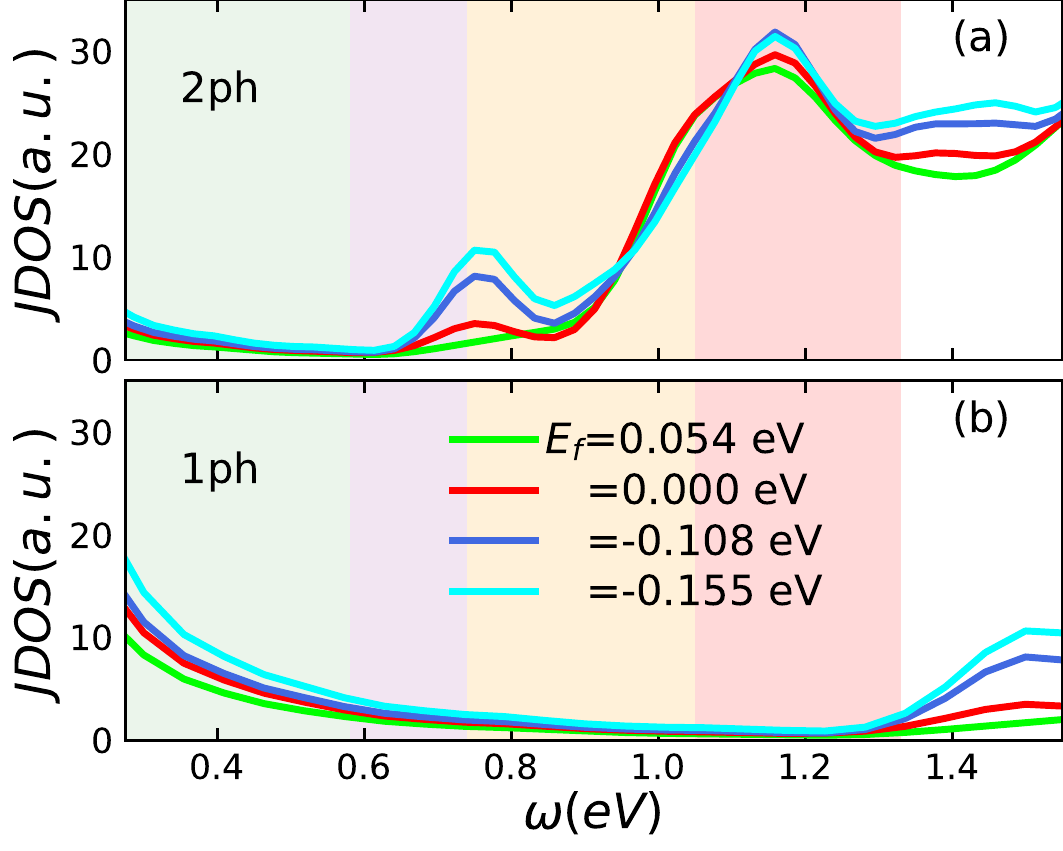}
    \caption{
    Optical joint density of states for SHG from two-photon~(a) and one-photon~(b) contributions. The parameters and colors are the same as Fig.~\ref{fig:experiment_vs_theory}.
    \label{fig:tjdos}}
\end{figure}

The small SHG yield in the green frequency window $\omega \in [0.276,0.58]$ in Fig.~\ref{fig:experiment_vs_theory}(c) is a result of the suppressed optical transitions between low-energy linearly dispersing bands close to the $\Gamma$ point; see green arrows in Fig.~\ref{fig:band}. To support this conclusion we first separate one-photon~($\omega$) and two-photon~($2\omega$) transitions contributing to $\chi^{xyz}$ in Fig.~\ref{fig:components_chi}. We observe that two-photon transitions dominate the green frequency region, regardless of whether the threefold $\Gamma$ node is occupied~($E_f=0.054$~eV) or unoccupied~($E_f=-0.155$~eV). Next, we compare this result to the two-photon and one-photon joint density of states~(JDOS) in Fig.~\ref{fig:tjdos}~(a) and~(b), respectively. The JDOS counts allowed optical transitions between occupied~(with energy $\omega_{n}$) and unoccupied~(with energy $\omega_{m}$) states ignoring their associated matrix elements, i.e.,  $\mathrm{JDOS}(\Omega) = \sum_{m,n} \delta(\omega_{m}-\omega_{n}-\Omega)$, where $\Omega=2\omega~(=\omega)$ for the two-(one-)photon JDOS. In the green frequency window, the one-photon JDOS dominates, compared to the two-photon JDOS, cf., Fig.~\ref{fig:tjdos}(a) and (b). Comparing with Fig.~\ref{fig:components_chi}, this indicates that the optical matrix elements suppress the one-photon contribution to $\chi^{xyz}$, reducing the overall SHG for $\omega <0.58$~eV. The band structure in Fig.~\ref{fig:band} suggests that the contribution to one-photon processes in this frequency region arises from linear bands around $\Gamma$, whose matrix elements therefore suppress SHG.

In order to understand further the low energy region and to benchmark our DFT calculations, we have developed a low-energy $k\cdot p$ model, see Supplemental Material~\cite{SuppMat}. 
This model captures low-energy excitations around the $\Gamma$ point, and brings insight into understanding the optical transitions resulting from the threefold node.
Specifically, the SHG response around $\Gamma$ displays a broad, low energy peak below the experimentally accessible frequencies. 
This single peak results from the merging of a dominant two-photon peak with a sub-dominant one-photon peak due to the large $\delta~\approx 100$~meV. A qualitatively similar broadened peak is also present by DFT when $\Delta=0$, which additionally receives contributions not captured by the $k\cdot p$ model. Consequently, the DFT peak is broader compared to that found using the $k\cdot p$ model. When $\Delta\neq 0$, the DFT results show that this low energy peak is largely suppressed. This is because $\Delta$ is a correction that pushes occupied and empty bands away from each other, and thus the optical transitions responsible for the peak are pushed to higher energies, resulting in a better agreement with the experimental data. These results highlight the fact that both the many-body corrections, modelled with a scissors potential $\Delta$, and the quasiparticle broadening $\delta$ are important to explain the experimental measurements. 

In addition, we note that the scissors potential $\Delta$ favors two-photon contributions. The reason is that, by separating occupied and unoccupied states, $\Delta$ reduces the available phase space for one-photon transitions with $\omega<\Delta$. In contrast, the phase space for two-photon transitions is only reduced for lower photon energies, $\omega< \Delta/2$. As a result, two-photon transitions dominate for $\omega<\Delta$, as seen in Fig.~\ref{fig:components_chi}.

We move on to analyze the purple frequency window in Fig.~\ref{fig:experiment_vs_theory}(c), i.e., $\omega \in [0.58,0.74]$~eV. The SHG increases in this region, a feature which is captured in our calculation only if $E_f<0$~eV. Separately plotting one- and two-photon contributions as before in Fig.~\ref{fig:components_chi} reveals that the two-photon response in the purple energy window is dominant. Consistent with our discussion in the previous paragraph, the rise of the two-photon contributions occurs around $\omega \approx \Delta/2$ in the JDOS. When compared to the band structure, the observation of a dominant two-photon transition in Fig.~\ref{fig:components_chi} suggests that two partially flat bands close to the $\Gamma$ point, separated by approximately $1.3$~eV, and connected by two-photon excitations~(purple arrows) in Fig.~\ref{fig:band}, are responsible for enhancing $\chi^{xyz}$ in the purple energy window. The width of this energy window is comparable to the quasiparticle broadening~($\delta =0.1$~eV), supporting their flat band origin.

At photon energies $\omega \in [0.74,1.05]$~eV, i.e., in the orange window in Fig.~\ref{fig:experiment_vs_theory}(c), the data exhibits a plateau-like structure. Our DFT calculations show that this feature is reproduced better for $E_f=-0.108,-0.155$~eV. Naively, one would expect that in this frequency window the one-photon electronic transitions from linear bands close to the $R$ point are activated. However, Figs.~\ref{fig:components_chi} and \ref{fig:tjdos} reveal that the one-photon contribution~(dashed lines) is small compared to the dominant two-photon transitions. The small contribution of the linearly dispersing bands close to the $R$ point is expected by dimensional analysis and confirmed by our results. The two-photon transitions responsible for SHG in this region likely involve dispersing valance bands around the $M$ and $\Gamma$ points, as indicated by the orange arrows in Fig.~\ref{fig:band}.

Lastly, there is a drastic increase of $\chi^{xyz}$ measured within the red energy window of $\omega \in [1.05,1.33]$~eV in Fig.~\ref{fig:experiment_vs_theory}(c). Our theoretical results also report an increased SHG yield in this energy range. Once more we can identify the substantial role of two-photon transitions compared to the smaller one-photon contribution, see Figs.~\ref{fig:components_chi} and \ref{fig:tjdos}. The large photon energies that define this energy window enable electrons to reach a considerable number of bands exemplified by red arrows in Fig.~\ref{fig:band}. As frequency increases, we observe quantitative differences between our DFT results and the experimental measurements, especially when $\omega > 1.33$~eV. These deviations could be attributed to the insufficient many-body corrections in our first principle calculations.

\paragraph*{\bf Conclusions-.}

In summary our SHG spectra on RhSi together with our first principles and $k\cdot p$ calculations show that one-photon transitions among relatively linear bands have a small contribution to SHG. Instead, two-photon transitions, including those between relatively flat bands, account for the observed SHG signal. At the single-particle level, this result is consistent with the expectation that one-photon transitions are more likely to connect linear bands close to $E_f$, which are expected by dimensional analysis to suppress the SHG. An additional, many-body effect, results from a sizable scissors potential $\Delta$, which separates occupied and unoccupied states and favours two-photon over one-photon transitions. Our DFT results indicate that similar observations apply to other monosilicides, like CoSi and MnSi~\cite{tangPRL2017,changPRL2017, raoNature2019, sanchez2019,takanePRL2019}. We expect that materials in the same space group , such as AlPt~\cite{schroterNatPhys2019}, PdGa~\cite{Schroter179,Sessi:2020dw}, and PtGa~\cite{Yao:2020cc} behave similarly.

Our findings complement earlier observations that predict the enhancement of SHG due to other factors, such as the skewness of the polarization distribution~\cite{Patankar2018} or a significant inter-site hopping~\cite{Tan2019}. Taken together, these results outline strategies to find materials with high SHG yield.

\paragraph*{\bf Acknowledgements-.}

S.~S. would like to thank G.~Davino for his suggestions on improving DFT calculations. We thank F.~de Juan, J.~E.~Moore, T.~Morimoto, J.~Orenstein, D.~Parker, L.~Wu, and Y.~Zhang for discussions and related collaborations. A.~G.~G. and S.~S. acknowledge funding by the ANR under the grant ANR-18-CE30-0001-01~(TOPODRIVE). A.~G.~G. is also funded by the European Union Horizon 2020 research and innovation program under grant agreement No.~829044~(SCHINES). M.~A.~S.~M. is supported by the European Union's Horizon 2020 research and innovation programme under the Marie-Sklodowska-Curie grant agreement No.~754303 and the GreQuE Cofund programme. D.~H.~T. acknowledges Temple University startup funding. K.M., and C.F. acknowledge the financial support from the European Research Council (ERC) Advanced Grant No. 742068 "TOP-MAT”; European Union’s Horizon 2020 research and innovation program (Grant Nos. 824123 and 766566) and Deutsche Forschungsgemeinschaft (DFG) through SFB 1143. K. M. acknowledges the Max Planck Society for the funding support under Max Planck–India partner group project.

\bibliography{SHG_RhSi.bib}

\newpage

\renewcommand{\thefigure}{S\arabic{figure}}
\setcounter{figure}{0} 

\appendix

\section*{Supplementary Materials}

\section{Experimental Details}

The light source was described in the main text as an optical parametric amplifier (OPA, Light Conversion - TOPAS Twins) as pumped by a regeneratively amplified Ti:sapph laser system (Coherent - Astrella). The polarization of the OPA output was purified using a linear wire grid polarizer (Thorlabs - WP12L-UB) and then passed through a quarter waveplate (Thorlabs - AQWP05M-980, AQWP05M-1600 or Alphalas - PO-TWP-L4-25-FIR) matched to the photon energy to produce a circularly polarized beam. After removing parasitic wavelengths due to other nonlinear optical processes in the OPA (as well as from interactions with the optics themselves), the beam was passed through a mechanically-driven polarizer spinning at 5~Hz in order to generate a varying incoming polarization angle $\phi$. For photon energies $> 0.480$~eV, the beam was then focused onto the sample using a 50~cm reflecting mirror at near-normal incidence so as to produce a relatively large spot. This enables high laser power to be incident on the sample while restricting the fluence to below the damage threshold. It also permitted for more SHG photons to be emitted per laser shot, yielding large enough signals to be measured by detection electronics in the IR frequency range where detector responsivity is relatively low. For photon energies $< 0.480$~eV, the beam instead was focused using a Cassegrain objective (Edmund Optics - 68-188) in order to obtain high enough fluences to produce measurable signals. The incidence angle introduced by the reflective objective was accounted for in the data analysis.

After reflecting from the sample, the beam was incident on a D-shaped mirror and then passed through an analyzer that was chosen to remain stationary in the vertical orientation to produce the signal $I(\phi)$. Upon emerging from the polarizer, the beam passed through a filter assembly to remove the fundamental wavelength while preserving the second-harmonic response. The filters that we used were: 2 shortpass 650~nm (Thorlabs - FESH0650) and 2 shortpass 700~nm filters for the $800 - 1200$~nm wavelength range; 2 shortpass 800~nm (Thorlabs - FESH0800) and 2 shortpass 1000~nm (Thorlabs - FESH0800) filters for the $1140-1500$~nm wavelength range; 2 longpass  600~nm (Thorlabs  - FELH0600), 2 shortpass 900~nm (Thorlabs - FESH0900) and 2 shorpass 1000~nm (Thorlabs - FESH1000) filters for the $1400-1620$~nm wavelength range; 2 longpass 700~nm  (Thorlabs - FELH0700) and 2 shortpass 1326~nm (Semrock - FF01-1326/SP-25) filters for the $1580-2000$~nm wavelength range; 2 shortpass 1326~nm (Semrock - FF01-1326/SP-25) and 2 shorpass 1550~nm (Spectrogon  - SP-1550) filters for the $2000-2600$~nm wavelength range; 2 shortpass 1550~nm (Spectrogon - SP-1550) for 2800~nm; and 2 shortpass 2600~nm filters (Spectrogon - SP-2600) for the $3500-4500$~nm wavelength range.

The detectors we used were: a multialkali photocathode photomultiplier tube (Hamamatsu - R12829) biased by a high voltage power
supply socket assembly (Hamamatsu - C12597-01) for incoming wavelength range $800 - 1620$~nm with transimpedance amplification performed by a charge sensitive preamplifier (Cremat CR-Z-PMT) in tandem with a shaping device (Cremat - CR-S-8us-US); an InGaAs photodiode (Thorlabs - FGA01) with transimpedance amplification performed by a charge sensitive preamplifier (Cremat - CR-Z-110) and a shaping device (Cremat - CR-S-8us-US) for incoming wavelength range $1580-2800$~nm; and a cooled InGaAs photodiode (Hamamatsu G12183-203K) for the $2600 - 4700$~nm incoming wavelength range, also attached to the same charge integrator/shaper as used for the $1580-2800$ photon range. In the $800-2800$~nm wavelength range, the intensity was recorded using a data acquisition card-based fast-sampling technique, a more detailed description of which is provided within Ref~\cite{Lu2019}, whereas for the $2600 - 4700$~nm wavelength range, the signal was measured using a lock-in amplifier (Zurich Instruments - MFLI) locked to the laser repetition rate.

Experiments were conducted on the polished (111) face of RhSi. Further details on the sample preparation can be found in a prior publication~\cite{rees2019}. On this face, the second-harmonic generation susceptibility tensor is given by
\begin{widetext}
\begin{equation}\label{eq:tensor}
\chi^{ijk}(2\omega;\omega,\omega) = \frac{1}{\sqrt{3}}\left(
\begin{array}{ccc}
 \left(\begin{array}{c}0\\ \sqrt{2}\chi^{xyz}\\ -\chi^{xyz}\end{array}\right) & \left(\begin{array}{c}\sqrt{2}\chi^{xyz}\\ 0\\ 0\end{array}\right) &
   \left(\begin{array}{c}-\chi^{xyz}\\ 0\\ 0\end{array}\right) \\
 \left(\begin{array}{c}\sqrt{2}\chi^{xyz}\\ 0\\ 0\end{array}\right) & \left(\begin{array}{c}0\\ -\sqrt{2}\chi^{xyz}\\ -\chi^{xyz}\end{array}\right) & \left(\begin{array}{c}0\\ -\chi^{xyz}\\ 0\end{array}\right) \\
 \left(\begin{array}{c}-\chi^{xyz}\\ 0\\ 0\end{array}\right) & \left(\begin{array}{c}0\\ -\chi^{xyz}\\ 0\end{array}\right) & \left(\begin{array}{c}0\\ 0\\ 2\chi^{xyz}\end{array}\right) \\
\end{array}
\right).
\end{equation}
\end{widetext}
The data were taken for incident polarization of the fundamental light dynamically rotating as angle $\phi$. As detailed above, there was a static polarizer in front of the detector to measure the emitted SHG for both vertical outgoing polarization, referred to as $I_{0^\circ}$. Using the tensor of Eq.~\eqref{eq:tensor}, we get
\begin{equation}\label{eq:PP}
I(\phi) = \frac{2}{3}\left[\chi^{xyz}(2\omega;\omega,\omega)\cos(2\phi)\right]^2
\end{equation}
as the expected response to which the data of Fig.~\ref{fig:experiment_vs_theory} were fit. In order to build a spectrum, these fits were also controlled for variable laser parameters including the incident power, spot size and pulse duration, as well as the detector responsivity as a function of measured photon energy. More details of this normalization process can be found in Ref.~\cite{Patankar2018}. We note that we did not need to account for the optical filter response since the exclusive use of long-pass and short-pass filters did not measurably affect the amplitude of the emitted SHG.

\section{Second-Harmonic Generation Response Function Within the Scissors Approximation}\label{Append:SHG-scissor}

\begin{figure}
    \centering
    \includegraphics[width=\columnwidth]{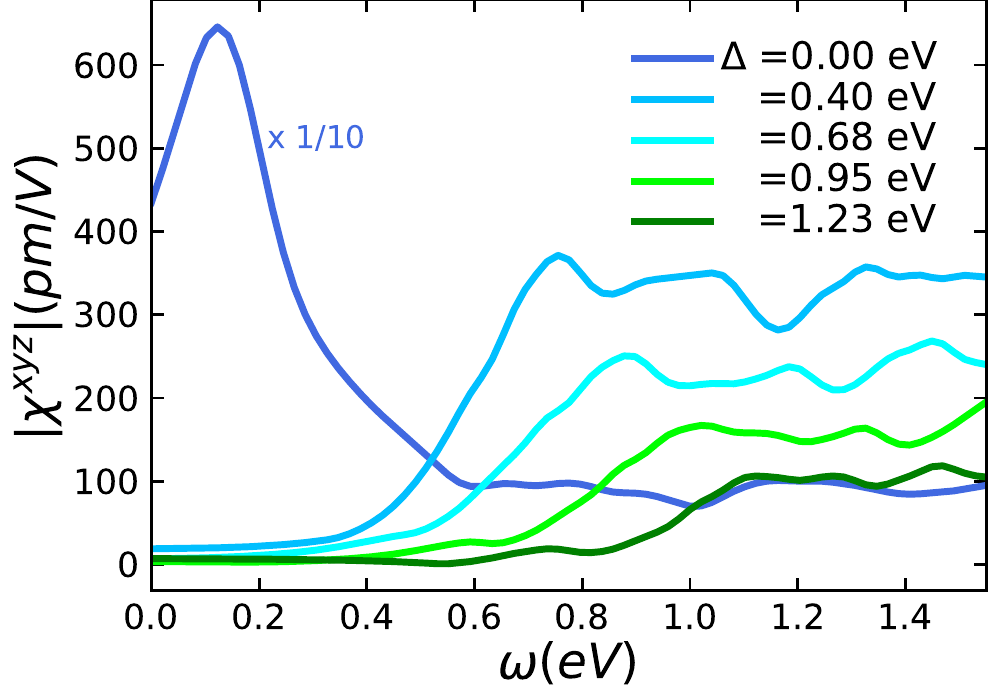}
    \caption{Different scissors corrections for $\chi$ in pristine RhSi ($E_f=0$) with a disorder broadening of $\delta=0.1$~eV.
    \label{fig:scissors}}
\end{figure}

The nonlinear polarization describing second-harmonic generation induced by an electric field $E^{b}(\omega)$ at a frequency $\omega$ along the Cartesian coordinate $b$ is written in the length gauge~\cite{Hughes1996} as
\begin{equation}
    P^{a} (2 \omega) = \chi^{abc} (2 \omega; \omega, \omega) 
    E^{b}(\omega) E^{c}(\omega),
\end{equation}
where $\chi^{abc}$ is the second-order susceptibility which satisfies the intrinsic permutation symmetry $\chi^{abc} =\chi^{acb}$.

The non-linear response function $\chi^{abc}$ accounts for interband and intraband contributions~\cite{Nastos2005}, and has the form   

\begin{align}
    \chi^{abc}(2\omega; \omega, \omega) 
    =&
     \chi_{\rm 2ph}^{abc} (\omega) + \chi_{\rm 1ph}^{abc} (\omega),
     \\
     =&
     \chi_{\rm 2ph,inter}^{abc} (\omega)
     +
     \chi_{\rm 1ph,inter}^{abc} (\omega)
     \nonumber\\
     &
     +
     \chi_{\rm 2ph,intra}^{abc} (\omega)
     +
     \chi_{\rm 1ph,intra}^{abc} (\omega)
     \nonumber\\
     &
     +
     \sigma^{abc} (\omega),
\end{align}
where the labels $\rm 2ph$ and $\rm 1ph$ denote two- and one-photon transitions, respectively, $\chi_{\rm 2ph}^{abc}=\chi_{\rm 2ph,inter}^{abc} +  \chi_{\rm 2ph,intra}^{abc}$, and $\chi_{\rm 1ph}^{abc}= \chi_{\rm 1ph,inter}^{abc} + \chi_{\rm 1ph,intra}^{abc} + \sigma^{abc}$. The above terms of $\chi^{abc}$ are 
\begin{align}
 \chi_{\rm 2ph,inter}^{abc} (\omega)
 =&
C \int_{k} \sum_{nml} \frac{ r^{a}_{nm} \{ r^{b}_{ml} r^{c}_{ln} \}
}{
(\omega_{ln} - \omega_{ml}) 
}
\frac{ 2 f_{nm}}{(\omega_{mn} - 2 \omega)}
\label{eq:chi2phinter}
,\\
 \chi_{\rm 1ph,inter}^{abc} (\omega)
 =&
 C \int_{k} \sum_{nml} \frac{ r^{a}_{nm} \{ r^{b}_{ml} r^{c}_{ln} \}
}{
(\omega_{ln} -  \omega_{ml})
}
\Big[
\frac{f_{ml}}{\omega_{ml} -\omega}
+\frac{f_{ln}}{\omega_{ln}-\omega}
\Big],
\label{eq:chi1phinter}
 \end{align}
 \begin{align}
 \chi_{\rm 2ph,intra}^{abc} (\omega)
 =&
C \int_{k} \sum_{nm} \frac{ r^{a}_{nm} \{ \Delta^{b}_{mn} r^{c}_{mn} \}
}{
\omega_{mn}^2  
}
\frac{ -8 {\rm i} f_{nm}}{(\omega_{mn} - 2 \omega)}
\nonumber \\
&+C \int_{k} \sum_{nml} \frac{ r^{a}_{nm} \{ r^{b}_{ml} r^{c}_{ln} \}
}{
\omega_{mn}^2 
}
\frac{ 2 f_{nm} (\omega_{ml} -\omega_{ln}) }{(\omega_{mn} - 2 \omega)}
\label{eq:chi2phintra}
,\\
 \chi_{\rm 1ph,intra}^{abc} (\omega)
 =&
 C \int_{k} \sum_{nml}r^{a}_{nm} \{ r^{b}_{ml} r^{c}_{ln} \}
\frac{ \omega_{mn} f_{nl}}{\omega^2_{ln} (\omega_{ln} -\omega)}
\nonumber \\
&-
 C \int_{k} \sum_{nml}r^{a}_{nm} \{ r^{b}_{ml} r^{c}_{ln} \}
\frac{ \omega_{mn} f_{ml}}{\omega^{2}_{ml}(\omega_{ml}-\omega)}
,
\label{eq:chi1phintra}
 \end{align}
 \begin{align}
    \sigma^{abc} (\omega)
    =&\frac{ {\rm i} C}{2}
    \int_{k}
    \sum_{nml}\omega_{nl} r^{a}_{lm} \{ r^{b}_{mn} r^{c}_{nl} \}
    \frac{f_{nm}}{\omega_{mn}^2 (\omega_{mn} -\omega)}
    \nonumber \\
    &-
    \frac{ {\rm i} C}{2}
    \int_{k}
    \sum_{nml}\omega_{lm} r^{a}_{nl} \{ r^{b}_{lm} r^{c}_{mn} \}
    \frac{f_{nm}}{\omega_{mn}^2 (\omega_{mn} -\omega)}
     \nonumber \\
     &+
     \frac{ {\rm i} C}{2}
    \int_{k}
    \sum_{nm}
    \frac{f_{nm} \Delta_{nm}^{a} \{r^{b}_{mn} r^{c}_{nm} \}  }{
    \omega_{mn}^2 (\omega_{mn} -\omega)
    },
\end{align}
where $C=e^3/\hbar^2$, the wave vector $k$ is defined in the Brillouin zone, $\int_k=\int {\rm d}^3k/(4\pi^3)$, lowercase Roman subscripts denote band indices, the energy of band $n$ is $\hbar \omega_{n}$, and the frequency difference is defined as $\omega_{mn}=\omega_{m}-\omega_{n}$. Here, $r_{mn}$ are matrix elements of the position operator given by $r^{a}_{nm}=v^{a}_{nm}/({\rm i} \omega_{mn})$, and $\Delta^{a}_{mn}= v^{a}_{mm}-v^{a}_{nn}$, where $v^{a}_{nm}$ denote the velocity matrix elements. The curly brackets impose symmetrization with respect to the Cartesian coordinates such that $\{ A_{ml}^a B_{ln}^{b} \} = \frac{1}{2} (A_{ml}^a B_{ln}^{b}   + B_{ml}^a A_{ln}^{b}) $.

\begin{figure}
    \centering
    \includegraphics[width=\columnwidth]{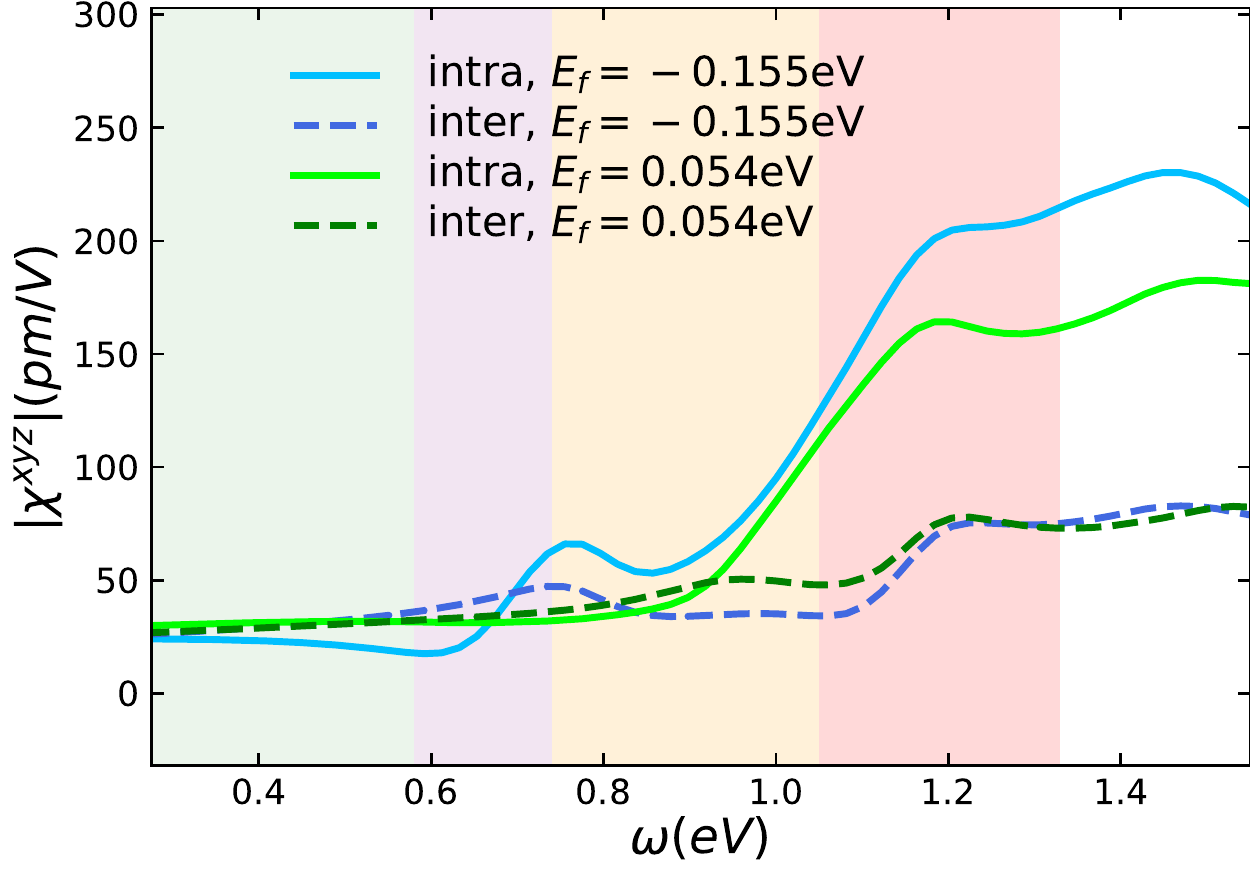}
    \caption{
    Intraband~(solid lines) and interband~(dashed lines) contributions of $\chi^{xyz}$ of RhSi at $E_{f}=-0.155$eV~(blue) and $E_{f}=0.054$eV~(green). The parameters and colors of shaded areas are the same as Fig.~1(c) in the main text.
    \label{fig:inter_intra}}
\end{figure}

To evaluate the matrix elements and band energies, we have calculated the ground-state properties using the density functional theory~(DFT). While these calculations can provide a satisfactory description for the occupied states, treating unoccupied states, which might be occupied during optical transitions, lacks many-body effects. As a result, the theoretical calculation of optical responses fails to exactly evaluate the energies at which the photons will be absorbed~\cite{Wang1981, Bergfeld2003, Yang2020}. There are two main approaches to remedy this mismatch between theoretical and experimental results: (i) the many-body GW formalism~\cite{Hybertsen1985, Aryasetiawan1998, Jiang2013}, and (ii) the scissors approximation~\cite{Levine1989, Hughes1996, Nastos2005, Sadhukhan2020, Song2020}.

The GW corrections are computed using a many-body self-energy. This self-energy corrects the energy gaps between occupied and unoccupied bands and thus improves the agreement between theoretical and experimental measurements~\cite{Zhandun2020}. Despite these advantages, converging the well-established GW self-consistent loop is computationally demanding. 

In this work we tackle the problem of inaccurate band gaps between occupied and unoccupied states using a scissors shift~($\Delta$). Within the scissors approximation, the position and velocity matrix elements are modified as
\begin{align}
    r_{nm} = \frac{
    \tilde{v}_{nm}
    }{
    {\rm i} \big[ \omega_{nm} +\frac{\Delta}{\hbar} (\delta_{n}^{\rm unoc} - \delta_{m}^{\rm unoc}) \big]
    },\\
    \tilde{v}_{nm} 
    =v_{nm}\frac{
    \omega_{nm} +\frac{\Delta}{\hbar} (\delta_{n}^{\rm unoc} - \delta_{m}^{\rm unoc})
    }{
    \omega_{nm}
    },
\end{align}
where $\delta_{n}^{\rm unoc}$ is the Kronecker delta for unoccupied state $n$. We illustrate the influence of $\Delta$ on shifting the absorption energy in the nonlinear response $\chi^{xyz}$ in Fig.~\ref{fig:scissors}. The results are calculated for the pristine RhSi system ($E_f=0$). The figure shows that by adjusting the energy gaps with $\Delta$, optical transitions can be modified.

\begin{figure}[t]
    \includegraphics[width=\columnwidth]{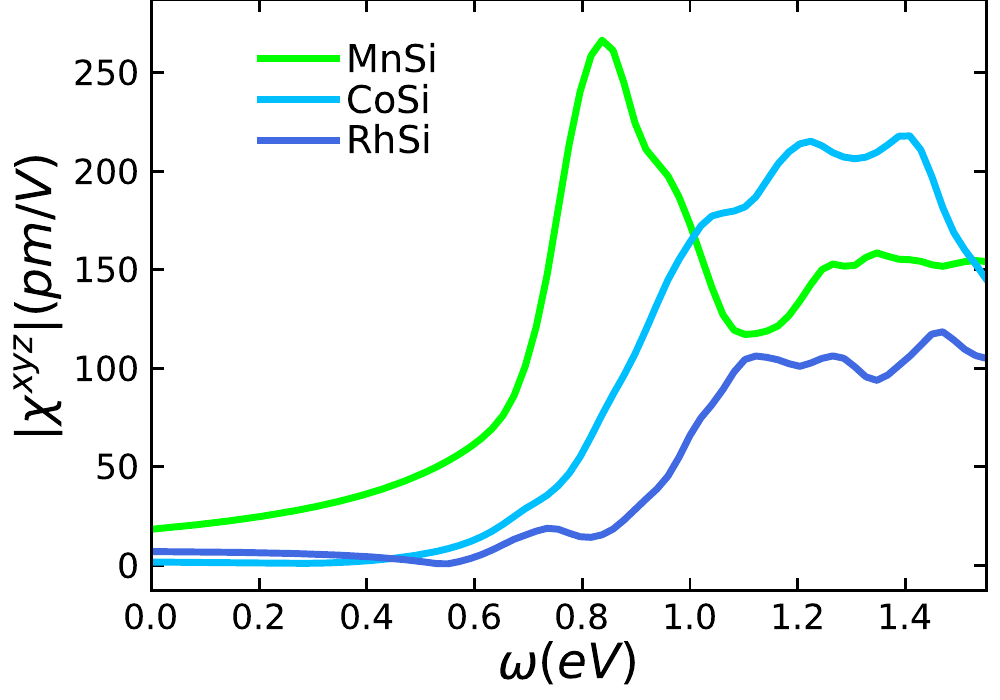}
    \caption{Comparison of $\chi^{xyz}$ with $\Delta=1.23$~eV, and $\delta=0.1$~eV for pristine MnSi, CoSi, and RhSi.
    \label{fig:SHG_family}}
\end{figure}

\section{Intraband and Interband Contributions in Second-Harmonic Generation of RhSi }

The intraband and interband contributions accounted in $\chi^{xyz}$ read
\begin{align}
    \chi^{xyz}_{\rm intra}(\omega) &= \chi^{xyz}_{\rm 2ph, intra}(\omega) + \chi^{xyz}_{\rm 1ph, intra}(\omega) ,\\
    \chi^{xyz}_{\rm inter}(\omega) &= \chi^{xyz}_{\rm 2ph, inter}(\omega) + \chi^{xyz}_{\rm 1ph, inter}(\omega),
\end{align}
where $\chi^{xyz}_{\rm 2ph, inter}$, $\chi^{xyz}_{\rm 1ph, inter}$, $\chi^{xyz}_{\rm 2ph, intra}$, and $\chi^{xyz}_{\rm 1ph, intra}$ are given by Eqs.~(\ref{eq:chi2phinter}, \ref{eq:chi1phinter}, \ref{eq:chi2phintra}, \ref{eq:chi1phintra}), respectively.
Fig.~\ref{fig:inter_intra} presents $\chi^{xyz}_{\rm intra}$ and $\chi^{xyz}_{\rm inter}$ for RhSi with $E_{f}=0.054, -0.155$~eV, see also Fig. \ref{fig:experiment_vs_theory}(c) for the total SHG yield. For $\omega  \lesssim 0.9$~eV the interband and intraband contributions exhibit comparable responses in the green, purple, and orange energy windows. Combined with the observation that two-photon transitions are dominant in RhSi, e.g., see Fig.~\ref{fig:components_chi} in the main text, the leading components responsible for $\chi^{xyz}$ in these regions are $\chi^{xyz}_{\rm 2ph, inter}$ and $\chi^{xyz}_{\rm 2ph, intra}$. For $\omega \in [1.05, 1.33]$~eV, the intraband contributions dominate compared with interband responses, and thus in this region the two-photon intraband transitions are responsible for the observed nonlinear SHG response.

\section{Second-Harmonic Generation in Family of Transition Metal Silicides}

\begin{figure}
    \centering
    \includegraphics[width=\columnwidth]{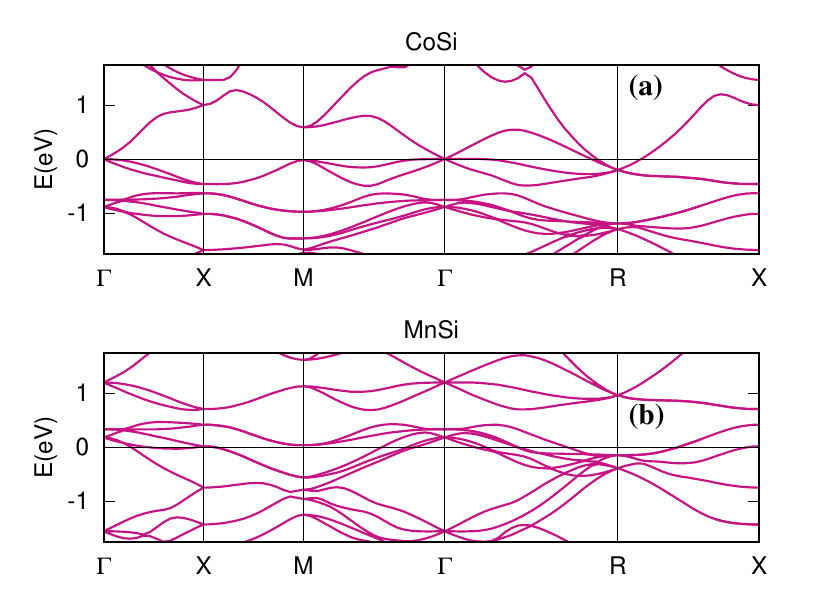}
    \caption{Electronic band structures of CoSi~(a) and MnSi~(b) along high-symmetry points of their Brillouin zone. Zero energies indicate the pristine Fermi energies. 
    \label{fig:band_mnsi_cosi}}
\end{figure}

To demonstrate the suppression of SHG in transition metal silicides, we have compared the nonlinear susceptibilities $\chi^{xyz}$ for the MnSi, CoSi and RhSi in Fig.~\ref{fig:SHG_family}. Our results confirm that this family of materials displays a comparable SHG. A larger SHG in MnSi can be attributed to a denser number of states at the Fermi level in comparison to CoSi and RhSi, cf. Figs.~\ref{fig:band}, and \ref{fig:band_mnsi_cosi}.

\section{Low-Energy Single-Particle Second-Harmonic Generation in RhSi: $k\cdot p$ model \label{sec:kp}}

To study the SHG at low energies near the $\Gamma$ point (shaded green region in Fig.~\ref{fig:experiment_vs_theory}) we use a three-band $k\cdot p$ model with up-to-second order terms in momentum. This model was originally presented in Ref.~\cite{Ni2020b} for CoSi, a material that crystallizes in the same space group (SG198) as RhSi. Here we give an overview of the construction of this model and the relevance of the different terms. For a more detailed explanation on the symmetries involved we refer the reader to Ref.~\cite{Ni2020b}.

It is illustrative to start by considering a higher-symmetry point group, $O$, to later on break the symmetry down to $T$, the physical point group at $\Gamma$, by including the necessary terms. 

We will work the Gell-Mann matrices $\lambda_{\alpha}$ which form a basis for the operators acting on the subspace of the three basis states for the threefold crossing at $\Gamma$ 

\begin{align}
\lambda_{0}=\id, \quad 
\lambda_{1}=
 \begin{pmatrix}
0  & -{\rm i} & 0 \\
{\rm i}& 0 & 0 \\
0 & 0 & 0
\end{pmatrix}, \quad 
\lambda_{2}=
 \begin{pmatrix}
0  & 0& -{\rm i} \\
0& 0 & 0 \\
{\rm i}& 0 & 0
\end{pmatrix},
\end{align}
\begin{align}
\lambda_{3}= 
\begin{pmatrix}
0  & 0 & 0 \\
0 & 0 & -{\rm i}\\
0 & {\rm i}& 0
\end{pmatrix}, \quad 
\lambda_{4}= 
\begin{pmatrix}
0  & 1 & 0 \\
1 & 0 & 0 \\
0 & 0 & 0
\end{pmatrix},
\end{align}
\begin{align}
\lambda_{5}= 
 \begin{pmatrix}
0  & 0 & 1 \\
0 & 0 & 0 \\
1 & 0 & 0
\end{pmatrix}, \quad 
\lambda_{6}=
 \begin{pmatrix}
0  & 0 & 0 \\
0 & 0 & 1 \\
0 & 1 & 0
\end{pmatrix},
\end{align}
\begin{align}
\lambda_{7}=
  \begin{pmatrix}
1 & 0 & 0 \\
0 & -1 & 0  \\
0 & 0 & 0
\end{pmatrix}, \quad 
\lambda_{8}= 
\begin{pmatrix}
\frac{1}{\sqrt{3}}  & 0 & 0 \\
0 & \frac{1}{\sqrt{3}} & 0 \\
0 & 0 & -\frac{2}{\sqrt{3}}
\end{pmatrix}. 
\end{align}

\begin{figure*}
    \centering
    \includegraphics[width=\textwidth]{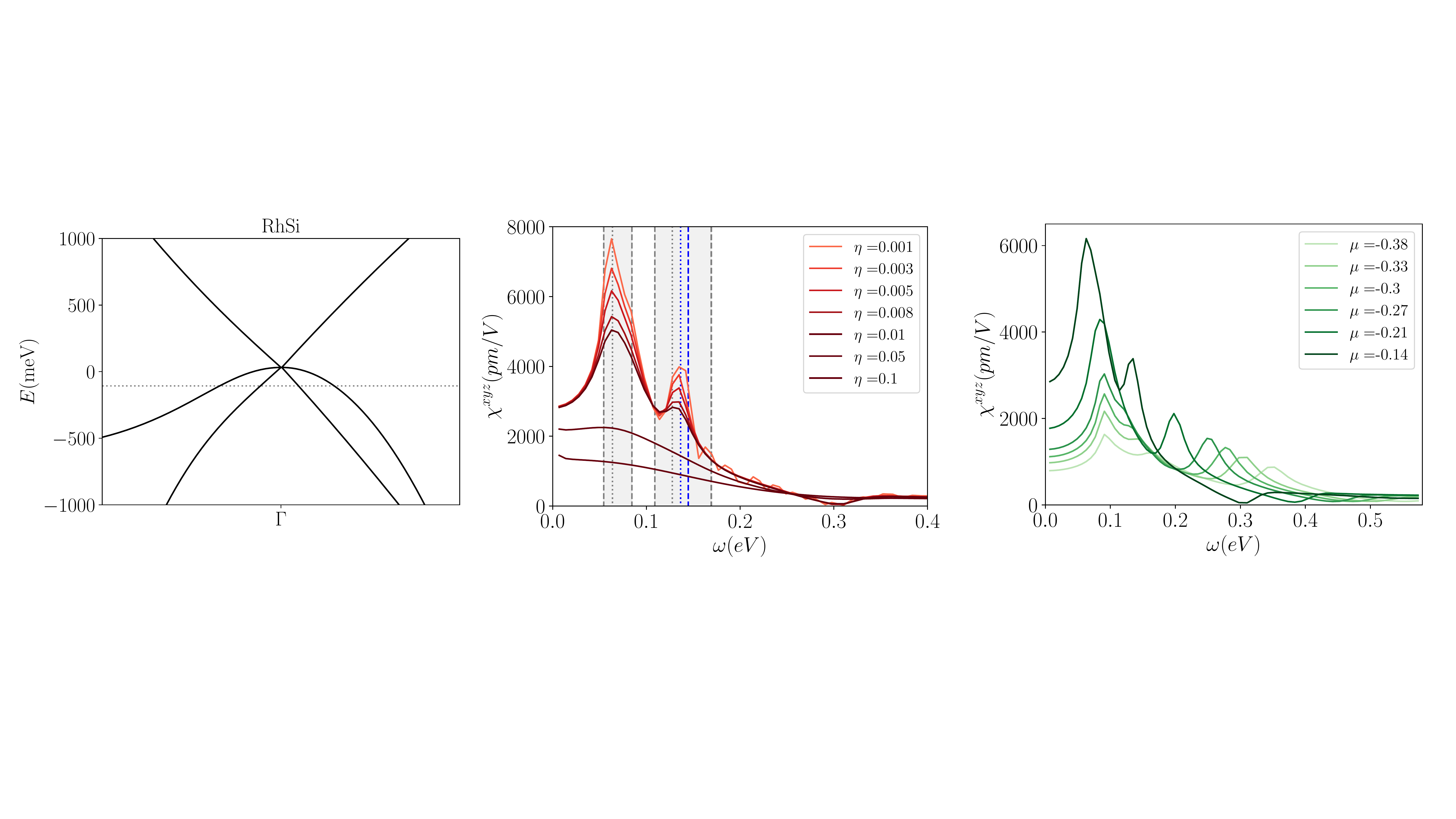}
    \caption{
        (a) Band structure of the $k\cdot p$ model along the $\Gamma\to X$ and $\Gamma\to M$ directions with parameters $(a,b,c,d,v_F)=(-0.0438344,-0.01,0.131377,0.1874,0.385)$. The threefold node is placed at $E_{\Gamma}=0$~eV, and the Fermi level is placed at $E_f=-0.14$~eV. The vertical dashed arrows indicate the most relevant one-photon activation energies.
        (b) Nonlinear susceptibility~(solid lines) of the $k\cdot p$ model  shown in (a) for different values of disorder $\delta$. The energy regions where the most relevant transitions are activated are indicated in shaded grey. The vertical dashed lines indicate the two-photon activation frequencies~(blue) and the one-photon activation frequencies~(ocher).
        (c) Nonlinear susceptibility for different values of $E_f$ shown in different shades of green with a disorder $\delta=0.005$ eV. 
        }
    \label{fig:shg_k_p_panel}
\end{figure*}

The point group $O$ is generated by $C_2$ rotations around (110), $C_2$ rotations around (100), and $C_3$ rotations around (111). This allows four different combinations of the Gell-Mann matrices that transform as the irreducible representations~(irreps) of $O$, 

\begin{align}
    A_1 &= \lambda_0,\\
    T_1 &= (-\lambda_2,\lambda_5,-\lambda_7),\\
    T_2 &= (\lambda_1, \lambda_4, \lambda_6),\\
    E   &= (-\frac{1}{2}\lambda_3+\frac{\sqrt{3}}{2}\lambda_8,-\frac{\sqrt{3}}{2}\lambda_3,-\frac{1}{2}\lambda_8),
\end{align}

where $A_1, T_1, T_2, E$ label the different irreps considered. We find for the same point group $O$ four momentum irreps up to second order in momentum, 
\begin{align}
    K_{A_1} &= k_x^2+k_y^2+k_z^2,\\
    K_{T_1} &= (k_x,k_y,k_z),\\
    K_{T_2} &= (k_y k_z, k_x k_z, k_x k_y),\\
    K_{E}  &= (k_x^2-k_y^2,(2k_z^2-k_x^2-k_y^2)/\sqrt{3}).
\end{align}

We can now build the most general symmetry-allowed Hamiltonian up to second order with point group symmetry $O$ by making scalar combinations of the momentum irreps with the Gell-Mann matrices, which reads
\begin{align}
\nonumber &H_O = \\ 
&\scalemath{0.9}{
    \begin{pmatrix}
    a k^2 +\frac{2c}{3}(k^2-3k_z^2) & {\rm i}v k_x + b k_y k_z & -{\rm i} v k_y + b k_x k_z \\
    -{\rm i} v k_x + b k_y k_z & a k^2 + \frac{2c}{3}(k^2 -3 k_y^2) & {\rm i} v k_z + b k_x k_y \\
    {\rm i} v k_y + b k_x k_z & -{\rm i} v k_z + b k_x k_y & a k^2 + \frac{2c}{3}(k^2 - 3 k_x^2)
    \end{pmatrix}
    },
\end{align}
where $k=\sqrt{k_x^2 + k_y^2 + k_z ^2}$, and $a$, $v_F$, $b$, and $c$ are the parameters corresponding to the terms coming from $A_1$, $T_1$, $T_2$, and $E$, respectively. 

Finally, we need to consider the point group $T$, obtained by breaking the $C_2$ rotations around (100) in the point group $O$. This leads to a new momentum irrep $K_{E}^{'}=(-(2k_z^2-k_x^2-k_y^2)/\sqrt{3}, k_x^2-k_y^2)$, allowing for a new term in the $k\cdot p$ Hamiltonian, which reads

\begin{align}
    H_T = H_O + \frac{2d}{\sqrt{3}}
    \begin{pmatrix}
    k_y^2-k_x^2 & 0 & 0 \\
    0 & k_x^2 - k_z^2 & 0 \\
    0 & 0 & k_z^2 - k_y ^2 
    \end{pmatrix}.
\end{align}

In previous calculations using this model (c.f. Ref~\cite{Ni2020b}) the effect of $d$ proved to be negligible in other optical responses like the circular photo-galvanic effect. Nevertheless, in this work, it is crucial to include the symmetry breaking term with a finite $d$ parameter since the SHG response is forbidden for the point group $O$, but generically finite for the point group $T$~\cite{Pozo2020,Ni2020b}.

The eigenvectors and eigenvalues of the $k\cdot p $ model do not depend on $d$ up to second order in $k$ \cite{Ni2020b,Pozo2020}, and thus we cannot use their analytical expressions to fit the value of $d$. To obtain the low-energy model parameters, we fit the existing four-band tight-binding model constructed for space group 198~\cite{changPRL2017,flickerPRB2018} to the DFT bands shown in Fig.~\ref{fig:band}, and obtain the $k\cdot p$ parameters by fitting the $k\cdot p$ model to the tight-binding model. The resulting values for the parameters are $(a,b,c,d,v_F)=(-0.0438344,-0.01,0.131377,0.1874,0.385)$. These parameters set the threefold node at $E=0$. Finally, to test the different values of $E_f$ and compare with the DFT calculation we add a term to the Hamiltonian $H=H_T-E_f\id_{3\times 3}$.

The nonlinear susceptibility of the $k\cdot p$ model features a two-peak structure at low energies. The energy regions where these peaks appear are delimited by the activation frequencies of the transitions from the lowest to the middle band and from the lowest to the upper band (see Fig.~\ref{fig:shg_k_p_panel}(a)). The first peak is dominated by the two-photon transitions from the lowest to the middle band (see Fig.~\ref{fig:shg_k_p_panel}(b), vertical dashed blue lines and shaded grey region). The lower, second peak in Fig.~\ref{fig:shg_k_p_panel}(b) appears in the energy region (shaded gray) delimited by one-photon transitions from the lower to the middle band (vertical ocher lines) and the two-photon transitions from the middle to the upper band (vertical blue lines).

As the disorder broadening $\delta$ is increased, the features of the nonlinear response are smoothed, and the two-peak structure is no longer distinguishable at $\delta=0.05$~eV (Fig.~\ref{fig:shg_k_p_panel}(b)) for $E_f=0.14$~eV. For higher $\delta$, the nonlinear response features a single, wider and smoother peak, similar to the one obtained in the DFT calculation with $\Delta=0$~eV (see Fig.~\ref{fig:scissors}~(a), dark blue curve).

The activation frequencies, and thus the position and width of the peaks, depend on the Fermi level. As the threefold node at $\Gamma$ is separated from the Fermi level, the activation frequencies and the difference between them become larger. As a result, the peak positions are shifted towards higher energies. The peaks also become wider, because the energy regions delimited by the activation frequencies are spread over a larger range of energies (see Fig.~\ref{fig:shg_k_p_panel}(c)). For large values of $E_f$ the one-photon transitions are suppressed due to the Pauli blocking at low energies, and the two-photon response becomes dominant.

\section{Comparing Low-Energy Second-Harmonic Generation using First Principles and $k\cdot p$ Calculations }

 \begin{figure}
     \centering
      \includegraphics[width=0.49\textwidth]{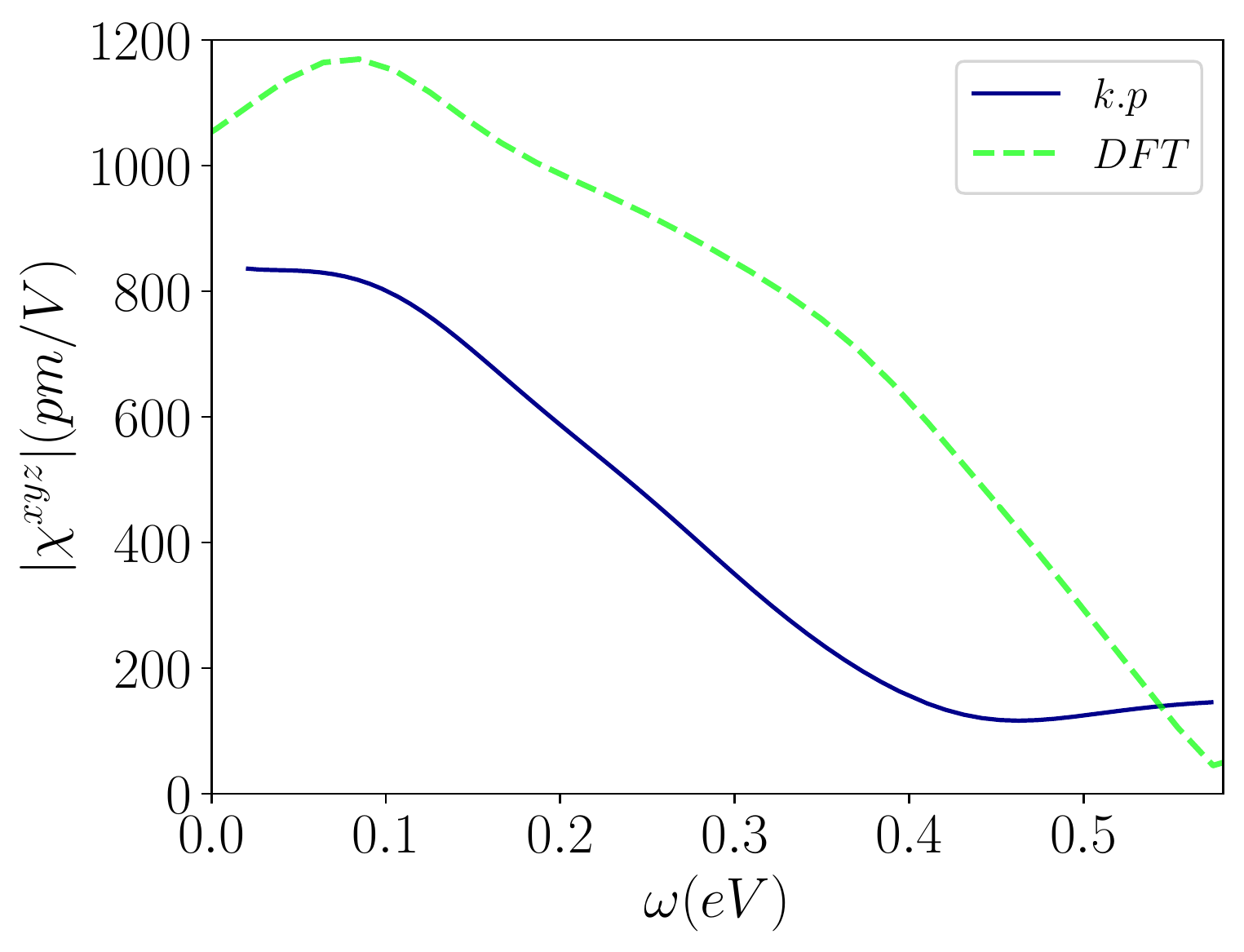}
     \caption{ Comparison between $\chi^{xyz}$ obtained within the $k\cdot p$ model~(blue line) at $E_f=-0.27$~eV and DFT calculations~(dashed green line) at $E_f = -0.14$~eV. Both curves are calculated with $\Delta=0$~eV and $\delta=0.1$~eV. 
     \label{fig:compare-dft-kp}
     }
 \end{figure}

To benchmark our DFT calculation we compare this calculation to our results obtained using the low-energy model around the $\Gamma$ point, described in Sec.~\ref{sec:kp}. Fig.~\ref{fig:compare-dft-kp} shows the nonlinear susceptibility of RhSi obtained for $E_f=-0.14$~eV using DFT and $E_f=-0.27$~eV using the $k \cdot p$ results, both computed without many-body effects, i.e., $\Delta =0 $~eV.  

Fig.~\ref{fig:compare-dft-kp} shows that the second-harmonic generation obtained from DFT and the $k \cdot p$ calculations are similar. The broader peak in DFT compared to that of the $k\cdot p$ model can be attributed to extra electronic transitions in DFT, which are not captured by the $k\cdot p$ model. As seen in Fig.~\ref{fig:scissors}, the scissors correction suppresses this peak and leads to a better description of the experimental SHG data in RhSi (Fig.~\ref{fig:experiment_vs_theory}).

\end{document}